\documentclass[%
 reprint,
nofootinbib,
 amsmath,amssymb,
 aps,
]{revtex4-1}

\usepackage{graphicx}
\usepackage{dcolumn}
\usepackage{bm}

\usepackage{multirow}
\usepackage{fancyhdr}
\usepackage{longtable}
\usepackage{booktabs}

\usepackage{units}
\usepackage{color}

\usepackage{lineno}

\begin{document}
\newcommand{\nuc}[2]{$^{#2}\rm #1$}

\newcommand{\bb}[1]{$\rm #1\nu \beta \beta$}
\newcommand{\bbm}[1]{$\rm #1\nu \beta^- \beta^-$}
\newcommand{\bbp}[1]{$\rm #1\nu \beta^+ \beta^+$}
\newcommand{\bbe}[1]{$\rm #1\nu \epsilon \epsilon$}
\newcommand{\bbep}[1]{$\rm #1\nu \epsilon \beta^+$}

\newcommand{\pic}[5]{
       \begin{figure}[ht]
       \begin{center}
       \includegraphics[width=#2\textwidth, keepaspectratio, #3]{#1}
       \end{center}
       \caption{#5}
       \label{#4}
       \end{figure}
}

\newcommand{\apic}[5]{
       \begin{figure}[H]
       \begin{center}
       \includegraphics[width=#2\textwidth, keepaspectratio, #3]{#1}
       \end{center}
       \caption{#5}
       \label{#4}
       \end{figure}
}

\newcommand{\sapic}[5]{
       \begin{figure}[P]
       \begin{center}
       \includegraphics[width=#2\textwidth, keepaspectratio, #3]{#1}
       \end{center}
       \caption{#5}
       \label{#4}
       \end{figure}
}

\newcommand{\picwrap}[9]{
       \begin{wrapfigure}{#5}{#6}
       \vspace{#7}
       \begin{center}
       \includegraphics[width=#2\textwidth, keepaspectratio, #3]{#1}
       \end{center}
       \caption{#9}
       \label{#4}
       \vspace{#8}
       \end{wrapfigure}
}

\newcommand{\baseT}[2]{\mbox{$#1\times10^{#2}$}}
\newcommand{\baseTsolo}[1]{$10^{#1}$}
\newcommand{\THL}{$T_{\nicefrac{1}{2}}$}

\newcommand{\UBI}{$\rm cts/(kg \cdot yr \cdot keV)$}

\newcommand{\Uflux}{$\rm m^{-2} s^{-1}$}
\newcommand{\Ucpd}{$\rm cts/(kg \cdot d)$}
\newcommand{\Uexpo}{$\rm kg \cdot d$}

\newcommand{\Qbb}{$\rm Q_{\beta\beta}\ $}

\newcommand{\validate}{\textcolor{blue}{\textit{(validate!!!)}}}

\newcommand{\improve}{\textcolor{blue}{\textit{(improve!!!)}}}

\newcommand{\missing}[1]{\textcolor{red}{\textbf{...!!!...} #1}\ }

\newcommand{\missref}{\textcolor{red}{[reference!!!]}\ }

\newcommand{\quanta}{\textcolor{red}{\textit{(quantitativ?) }}}

\newcommand{\misscite}{\textcolor{red}{[citation!!!]}}

\newcommand{\PC}{$N_{\rm peak}$}
\newcommand{\BIC}{$N_{\rm BI}$}
\newcommand{\PAPR}{$R_{\rm p/>p}$}

\newcommand{\PCR}{$R_{\rm peak}$}


\newcommand{\gline}{$\gamma$-line}
\newcommand{\glines}{$\gamma$-lines}
\newcommand{\gray}{$\gamma$-ray}
\newcommand{\grays}{$\gamma$-rays}


\newcommand{\tab}{Tab.~}
\newcommand{\eq}{Eq.~}
\newcommand{\fig}{Fig.~}
\renewcommand{\sec}{Sec.~}
\newcommand{\chap}{Chap.~}

 \newcommand{\fn}{\iffalse \fi} 
 \newcommand{\tx}{\iffalse \fi} 
 \newcommand{\txe}{\iffalse \fi} 
 \newcommand{\sr}{\iffalse \fi} 

\preprint{AIP/123-QED}

\title{Constraints on partial half-lives of $^{\bf 136}$Ce and $^{\bf 138}$Ce double electron captures}

\author{B. Lehnert}%
\email{bjoernlehnert@lbl.gov}
\affiliation{Nuclear Science Division, Lawrence Berkeley National Laboratory, Berkeley, CA 94720, U.S.A.}

\author{M. Hult}
\affiliation{European Commission, Joint Research Centre, 2440 Geel, Belgium}

\author{J. Kotila}
\affiliation{Finnish Institute for Educational Research, University of Jyv{\"a}skyl{\"a}, P.O. Box 35, FI-40014 Jyv\"askyl\"a, Finland}
\affiliation{Center for Theoretical Physics, Sloane Physics Laboratory, Yale University, New Haven, Connecticut 06520-8120, USA}

\author{G. Lutter}
\affiliation{European Commission, Joint Research Centre, 2440 Geel, Belgium}

\author{G. Marissens}
\affiliation{European Commission, Joint Research Centre, 2440 Geel, Belgium}

\author{A.~Oberstedt}
\affiliation{Extreme Light Infrastructure - Nuclear Physics (ELI-NP) / Horia Hulubei National Institute for Physics and Nuclear Engineering (IFIN-HH), 077125 Bucharest-Magurele, Romania}

\author{S.~Oberstedt}
\affiliation{European Commission, Joint Research Centre, 2440 Geel, Belgium}

\author{H. Stroh}
\affiliation{European Commission, Joint Research Centre, 2440 Geel, Belgium}

\author{K. Zuber}
\affiliation{Institute for Nuclear and Particle Physics, TU Dresden, 01069 Dresden, Germany}
\date{\today}

\begin{abstract}

The \gray\ emissions from a radiopure cerium-bromide crystal with a mass of 4381~g were measured for a total of 497.4~d by means of high-resolution \gray\ spectrometry in the HADES underground laboratory at a depth of 500 m.w.e. 
A search for \bbe{0/2} and \bbep{0/2} double beta decay transitions of $^{136}$Ce and $^{138}$Ce was performed using Bayesian analysis techniques. No signals were observed for a total of 35 investigated decay modes. 90\% credibility limits were set in the order of \baseTsolo{18-19}~a. 
Existing constraints from a cerium oxide powder measurement were tested with a different cerium compound and half-life limits could be improved for most of the decay modes. 
The most likely accessible decay mode of the $^{136}$Ce \bbe{2} transition into the $0^+_1$ state of \nuc {Ba}{136} results in a new best 90\% credibility limit of \baseT{5.0}{18}~a. 

\end{abstract}

\pacs{Valid PACS appear here}
\maketitle

\section{Introduction}\label{sec:intro}

Neutrinoless double beta (\bb{0}) decay is a second-order weak nuclear decay process requiring physics beyond the Standard Model (SM) of particle physics. 
The observation of $0\nu\beta\beta$ decay would establish lepton number violation (LNV) and neutrinos as Majorana particles. It would also open up possibilities to explain the matter-antimatter asymmetry in the Univserse converting Leptogenesis into Baryogenesis (see e.g.\ Ref.~\cite{dep18}).  
The inverse of the $0\nu\beta\beta$ decay half-life $T^{0\nu}_{1/2}$ in a given isotope is conventionally expressed as
\begin{equation}
\label{eq:halflife-intro}
	\left[T^{0\nu}_{1/2}\right]^{-1} = \left|f(m_i, U_{ei})\right|^2 \cdot G_{0\nu}  \cdot \left|M_{0\nu}\right|^2,
\end{equation}
with the phase space factor $G$ and the nuclear matrix element $M$. 
The LNV mechanism $f(m_i, U_{ei})$ can have many origins. Under the assumption of light Majorana neutrino exchange, it is connected with  $m^{eff}_\nu$ and allows to constrain the absolute neutrino mass scale.
While \bb{0} decay is one of the most promising avenues to search for new physics, the similar neutrino-accompanied double beta (\bb{2}) decay is allowed within the SM:  
\begin{equation}
2\nu\beta^-\beta^-: (Z, A) \longrightarrow (Z + 2, A) + 2 \, e^- + 2 \, \overline{\nu}_e\ ,
\end{equation}
with half-life
\begin{equation}
\label{eq:halflife-intro}
	\left[T^{2\nu}_{1/2}\right]^{-1} =  G_{2\nu} \cdot \left| M_{2\nu} \right|^2.
\end{equation}

It was detected directly in 11 isotopes so far with half-lives between 10$^{18}$ and 10$^{21}$~a~
\cite{tre02,sak13} and provides valuable experimental information to better understand the complex nuclear physics of double beta decay processes and isotopes.
Equivalent processes based on the right side of the mass parabola are decays such as (1) double electron capture $\epsilon\epsilon$, (2)  electron capture with positron emission $\epsilon\beta^+$ and (3) double positron emission $\beta^+\beta^+$:
\begin{eqnarray}
2\nu {\rm \epsilon\epsilon} :  2 \, e^- + (Z, A) & \longrightarrow & (Z - 2, A)  + 2 \, \nu_e\\
2\nu {\rm \epsilon} \beta^+ :     \, e^- + (Z, A) & \longrightarrow & (Z - 2, A)  +  e^+  + 2 \, \nu_e\\
2\nu \beta^+ \beta^+ :                  (Z, A) & \longrightarrow & (Z - 2, A)  + 2 \, e^+  + 2 \, \nu_e
\end{eqnarray}
The experimental sensitivity for those processes is considerably lower compared to $2\nu\beta^-\beta^-$ decay and, except for two cases, only lower limits for the half-lives in the range of \baseTsolo{18-21}~a could be obtained for a few isotopes~\cite{tre02,tre95,gav06,bel09,ruk10}. In Ref.~\cite{mesh01} the total weak decay ($\beta^+\beta^+$ + $\epsilon\epsilon$ + $\epsilon\beta^+$) half-life for $^{130}$Ba is reported as T$_{1/2}$ = (2.2 $\pm$ 0.5) $\times$ 10$^{21}$ a (68\% C.L.) based on a geochemical analysis of natural barite (BaSO$_4$).
Recently a first direct observation of \bbe{2} could be achieved in \nuc{Xe}{124} with a half-life of \baseT{1.8}{22}~a \cite{apr19}.\\

\bb{2} decays into excited states, the focus of this work, are so far only observed in two nuclides, \nuc{Nd}{150} and \nuc{Mo}{100}, with 
average half-lives of \baseT{1.33^{+0.45}_{-0.26}}{20}~a and \baseT{5.9^{+0.8}_{-0.6}}{20}~a, respectively \cite{ESAverage}. 
The first observations of these decay modes were performed with samples on HPGe detectors in the ``source$\neq$detector" approach that is also employed here. \\

Neutrinoless double-beta decay has not been observed, yet. Although much effort is being made to improve the sensitivity by using the  ``source$=$detector" approach, increasing the target mass, isotopic fraction of the nuclide under consideration and improving the radio-purity of the sample, only lower limits of the decay half-lives are reported. Leading half-life limits and sensitivities recently exceeded \baseTsolo{26}~a by the KamLAND-ZEN \cite{KLZ0nbb} and GERDA \cite{Gerda0nbb} experiments for \nuc{Xe}{136} and \nuc{Ge}{76}, respectively.\\

The recent development of production of large cerium-bromide detectors put cerium isotopes in the focus of double beta decay searches. 
Cerium has three isotopes, which are candidates for double beta decay: $^{136,138}$Ce and $^{142}$Ce. 
Moreover, being used as \gray\ detectors offers the possibility to exploit the ``source$=$detector" approach for cerium~\cite{bel03,ber97,bel11}. \\

In this work, we measure double beta decays in a cylindrical CeBr$_3$ crystal of size 10.2 cm $\times$ 10.2 cm (diameter $\times$ length) with 4381~g mass 
with an ultra-low-background HPGe detector in the ``source$\neq$detector" approach. 
The large size of the crystal contains a significantly greater number of cerium atoms than were available in previous searches. 
Thorough characterization of the material itself and other crystals produced by the same company revealed a very high radio-purity~\cite{lutter13,bil11}. This provides excellent background conditions for investigating rare decay processes and the possibility for future ``source$=$detector" measurements if additional background from the light readout is kept low. 
%

\section{Double beta decays in cerium}\label{sec:cerium}

Natural cerium consists of 4 isotopes out of which only \nuc{Ce}{140} (88.45\% abundance) is expected to be stable. The long lived isotopes \nuc{Ce}{142}, \nuc{Ce}{138}, and \nuc{Ce}{136} are candidates for double beta decay:

\begin{figure*}[ht]
\centerline{\includegraphics[width=0.9\textwidth]{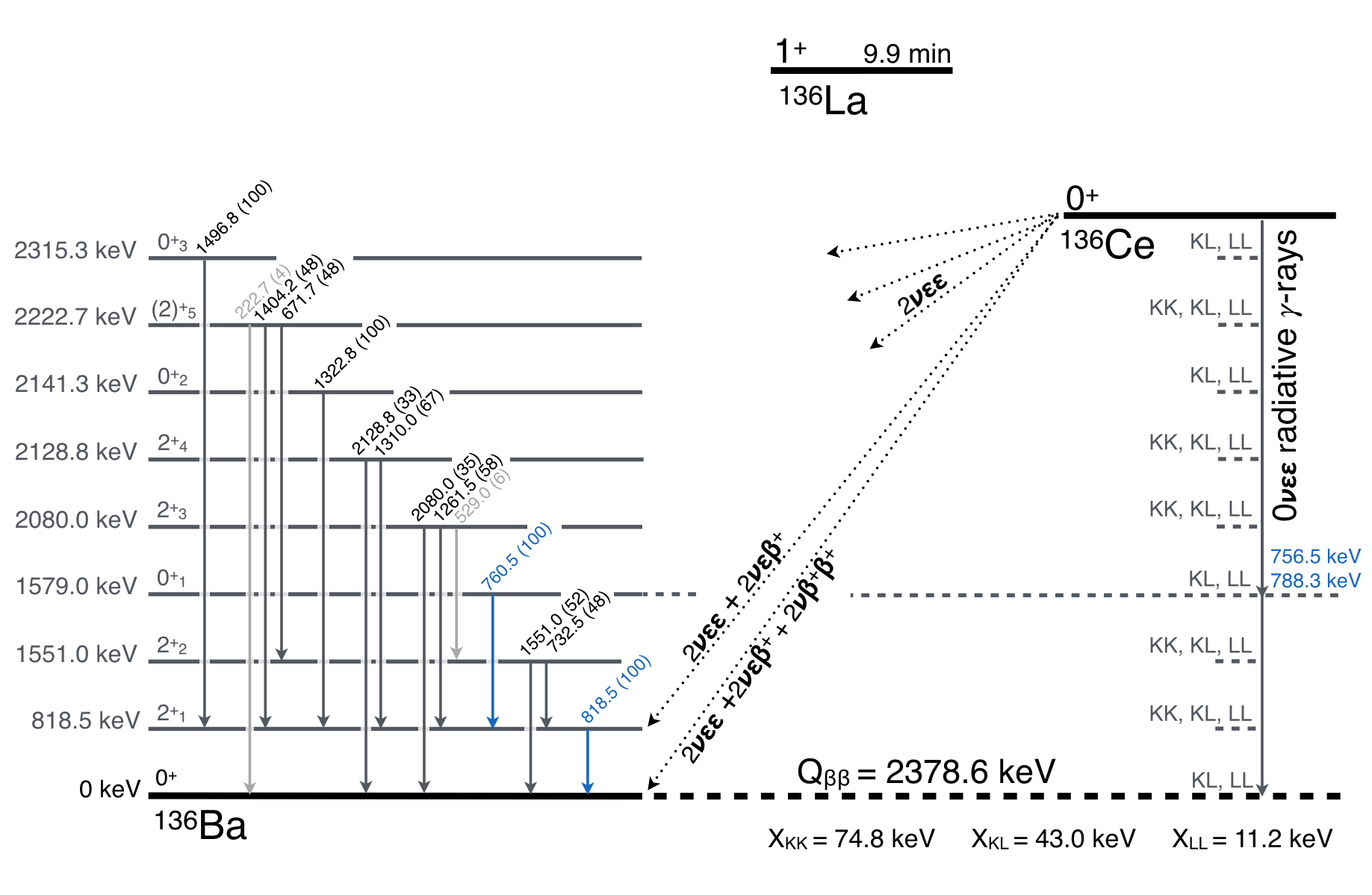}}
\vspace{-2mm}
\caption{Decay scheme of \nuc{Ce}{136}  to \nuc{Ba}{136}. 
Each excited state of \nuc{Ba}{136} shows the de-excitation cascade including \gray\ emission probabilities. 
Decay branches not considered in the analysis are shown in gray. All other \grays\ are included in the combined fits. 
The radiative \bbe{0} decay is illustrated on the right for the single bremsstrahlung \gray\ emission case. 
\bbe{0} to a final 2$^+$ state can occur with captures from any combination of the K and L shell, where each shell electron reduces the available energy for the bremsstrahlung \gray. 
For \bbe{0} to a final 0$^+$ state only KL and LL captures are allowed due to spin constraints.  
The highlighted example in blue shows the \bbe{0} decay modes into the first excited 0$^+$ and its subsequent \gray\ de-excitation. 
\bbep{0/2} and \bbp{0/2} decay can only occur into the ground and first excited state. All other excited states are populated by \bbe{0/2}.
\label{fig:Ce-136-decay}}
\end{figure*}

\begin{itemize}
\item \nuc{Ce}{142} has a high natural abundance of 11.11\% and is a candidate for the \bbm{2} decay with an energy release of $Q = (1416.8\pm 2.9)$~keV \cite{Qvalues}. 
The first excited state is at 1575.8~keV which is higher than the Q-value. Hence, no excited state of its daughter $^{142}$Nd can be populated and the decay cannot be measured by \gray\ spectrometry.

\item \nuc{Ce}{138} has a natural abundance of 0.251\% and is a candidate for the \bbe{2} decay with $Q = (690.6 \pm 5.0)$~keV \cite{Qvalues}. 
Also in this case no excited state of \nuc{Ba}{138} can be populated and the experimental signature for the $2\nu$ mode are only barium X-rays with energies below 40~keV. Due to the high absorption of the low energy X-rays, this decay channel will not be studied. 
In case of the $0\nu$ mode, the excess of energy due to the non-emission of the 2 neutrinos can be released through a bremsstrahlung photon in addition to the X-rays. The energy of the emitted photon, E$_{0\nu}$, is equal to:
\begin{eqnarray}
E_{0\nu} = Q - E_{\beta1} - E_{\beta 2} \label{qe:0nECEC}
\end{eqnarray}

where E$_{\beta 1}$ and E$_{\beta 2}$ are the binding energies of the K, L1, L2, L3 shells. 
In case of barium these are E$_K$ = 37.4~keV, E$_{L1}$ = 6.0~keV, E$_{L2}$ = 5.6~keV and E$_{L3}$ = 5.2~keV.
Due to  the energy resolution of the \gray\ spectrometer the E$_L$ energies cannot be distinguished. Therefore, the mean value of E$_L$, 5.6~keV, is taken for the analysis.

For the $0\nu$LK, and $0\nu$LL decays of $^{138}$Ce we expect the emission of a photon at (647.6 $\pm$ 5) keV and (679.4 $\pm$ 5)~keV. Note that due to angular momentum constraints the $0\nu$KK decay to a $0^+$ state is not possible \cite{doi93}.

\item \nuc{Ce}{136} has a natural abundance of 0.185\% and is a candidate for \bbe{2}, \bbep{2}, and \bbp{2} decay with $Q = (2378.53 \pm 0.27)$ keV \cite{QValueCe136}. Double beta decays of this isotope are the main focus of this work and the large number of possible decay modes are illustrated in \fig \ref{fig:Ce-136-decay}. 
The \bbp{2} mode can only populate the ground state of \nuc{Ba}{136} since each $\beta^+$ reduces the Q-value twice by 511~keV.  
Consequently, the \bbep{2} mode can populate the ground state and first excited $2^+_1$ state at 818.5~keV. 
The \bbe{2} decay mode can populate a total of 8 excited states in \nuc{Ba}{136} as shown in \fig \ref{fig:Ce-136-decay}. The most likely mode among the excited state transitions investigated here is the first excited $0^+_1$ state at 1579.0~keV.
The $0\nu$ modes are possible into the ground and excited states with additional emission of a radiative bremsstrahlung photon to release the excess energy as shown in \eq \ref{qe:0nECEC}. This is illustrated in \fig \ref{fig:Ce-136-decay} on the right. For each final state, the bremsstrahlung energy depends on the shell combination from which the two electrons are captured. For the $2^+$ final states, KK, KL, and LL combinations are considered, each differing in the energy $E_{0\nu}$ but with an otherwise identical \nuc{Ba}{136} \gray\ de-excitation cascade. For the $0^+$ states only the KL and LL shell captures are considered. Figure~\ref{fig:Ce-136-decay} shows the example of bremsstrahlung energies for the \bbe{0} $0^+_1$ state transitions in blue.

\end{itemize}

In \nuc{Ce}{136} a potential resonant double-electron capture to the excited levels in \nuc{Ba}{136} with energies of 2392.1~keV and 2399.9~keV could reduce the half-life by several orders of magnitude to the order of \baseTsolo{24}~a (e.g.~\cite{bern83}). However, a later measurement determined the reaction Q-value of this process to Q = (2378.53 $\pm$ 0.27) keV \cite{QValueCe136} ruling out such a resonance enhancement in the \nuc{Ce}{136} - \nuc{Ba}{136} system.\\

Table~\ref{tab1} summarizes the fundamentally different decay modes for each cerium isotope of interest. 
Each of these modes has a number of possible ``sub-modes" either going to different excited states or starting from captures on different electronic shell combinations (see \fig \ref{fig:Ce-136-decay}). The number of sub-modes investigated in this work are listed in the second column. 
In this manuscript, the \bbe{2} transition into the $0^+_1$ with the 818.5~keV and 760.5~keV \gray\ is used to exemplarily illustrate the analysis for all other decay modes.\\

A clear signature of the 511~keV annihilation peak in combination with the 510.8~keV \gline\ from \nuc{Tl}{208} is present in the background data (as shown later in \fig \ref{fig:SpectrumWide}). Hence, the search for decay modes including a $\beta^+$ cannot reliably use this signature and we only constrain the \bbep{0/2} mode into the $2^+_1$ state with the de-excitation \gline\ at 818.5~keV. Due to the possible coincidences of the de-excitation \gray\ with the annihilation \grays, this mode has a slightly reduced detection efficiency compared to the pure de-excitation \gray\ search of the \bbe{0/2} into the same final state. Also note that the signature for the \bbep{0} and \bbep{2} modes are identical since the $\beta^+$ can carry the remaining decay energy and no radiative bremsstrahlung emission is expected for the $0\nu$ case with a $\beta^+$ emission.\\

The right column in \tab \ref{tab1} lists corresponding theoretical half-life estimates when available. The theoretical half-lives are given for the ground state (g.s.) and first excited $0^+_1$, as well as first and second  $2^+$ state transitions when applicable. For completeness, \tab \ref{tab1} also lists theoretical results for \nuc{Ce}{142}, which are however not investigated in this work.\\

\begin{table*}[ht]
\caption{Summary of the double beta decay isotopes in cerium together with their theoretical half-life estimates. 
Listed are the main $2\nu$ and $0\nu$ decay modes in the first column and the number of sub-modes available due to various excited final states or due to different starting energies based on captures from different shell combinations in the second column. The number of sub-modes investigated in this work are shown in parenthesis.  
Half-life predictions in the literature are shown in the third column for ground state transitions, as well as excited state transitions where available. For the $0\nu$ modes, the values are given for $m_\nu^{\rm eff} = 1$~eV.
The theoretical expectations marked as IBM-2 are calculated with matrix elements taken from/calculated as described in Ref.~\cite{CeNME}  and phase space factors taken from/calculated as described in Refs.~\cite{Kot12, CePSF}.
}\label{tab1}
\begin{center}
\begin{tabular}{lcl}
\\[-2mm]
\hline
\\[-3mm]
     decay mode  &  sub modes    &theoretical T$_{1/2}$\\
                          &  all (investigated) &   [a] \\[2mm]
\hline\\[-3mm]
$^{136}$Ce      	\\					
  \bbe{2}   & 9 (8)    &\underline{g.s}:  \baseT{1.2}{21} [IBM-2],  \baseT{1.7}{22} \cite{hir94}, \baseT{9.6}{21} \cite{rum98},        \\     
  		&        & \baseT{0.3-6.4}{19} \cite{suh93}, \baseT{3.2-5.1}{21} \cite{civ98},  \baseT{3.7}{23} \cite{aba84}, \\
		&        & \baseT{1.6-5.9}{22} \cite{pir15}\\
		& &\underline{$2^+_1$}:  \baseT{2.3-8.5}{29} \cite{pir15}\\
		& &\underline{$0^+_1$}:  \baseT{5.7}{25} [IBM-2], \baseT{7.7-28}{29} \cite{pir15}\\
		& &\underline{$2^+_2$}:  \baseT{7.3-27}{33} \cite{pir15}\\
		
  \bbep{2} &  2 (1)    &\underline{g.s}:  \baseT{7.8}{22} [IBM-2], \baseT{9.2}{23} \cite{hir94},         \baseT{6.0}{23}  \cite{rum98}, \\
   && \baseT{2.8}{24} \cite{aba84}, \baseT{1.0-3.7}{24} \cite{pir15}\\
  		& &\underline{$2^+_1$}:  \baseT{7.8-29}{30} \cite{pir15}\\

  \bbp{2}    & 1 (0)   &\underline{g.s}:  \baseT{5.2}{30} [IBM-2], \baseT{5.2}{31} \cite{hir94},  \baseT{9.6}{31} \cite{aba84}, \\
  		 &             & \baseT{6.8-25}{31} \cite{pir15}\\
  
  \bbe{0}    &  23 (23) & -\\
  \bbep{0} & 2 (1) &\underline{g.s}:  \baseT{3.3}{26} [IBM-2], \baseT{1.8}{26} \cite{suh03}, \baseT{6.4-110}{25} \cite{suh98}    \\
  \bbp{0}   & 1 (0) &\underline{g.s}: \baseT{2.0}{29} [IBM-2], \baseT{3.8}{30} \cite{suh03}, \baseT{5.6-7.3}{29} \cite{suh98}\\
\\[-2mm]\hline\\[-3mm]
$^{138}$Ce  \\					      
  \bbe{2}   & 2 (0) &\underline{g.s}:  \baseT{1.5}{24} [IBM-2],\baseT{2.1}{26} \cite{aba84} \\
  \bbe{0}   &  2 (2) & -\\
\\[-2mm]\hline\\[-3mm]
$^{142}$Ce \\					     
  \bbm{2}  & 1 (0)  &\underline{g.s}:  \baseT{2.5}{22} [IBM-2], \baseT{1.6}{21} \cite{sta90},   \baseT{0.16 - 23}{22} \cite{bob04}, \\
                 &          & \baseT{3.8}{22} \cite{del17}\\
  \bbm{0}  & 1 (0) &\underline{g.s}:  \baseT{2.1}{25} [IBM-2], \baseT{2.8}{24} \cite{sta90}  \\
\\[-2mm]
\hline
\end{tabular}
\end{center}
\end{table*}

For this work, the half-lives labeled IBM-2 (microscopic Interacting Boson Model) in \tab \ref{tab1} were obtained by dedicated calculations of nuclear matrix elements as described in Ref.~\cite{CeNME} and phase space factors (PSF) as described in Ref.~\cite{Kot12} and \cite{CePSF}. In the calculation of PSFs, the same binding energies for K and L shells were used as in the following analysis, i.e, E$_K$ = 37.4~keV and the mean value, 5.6~keV for E$_L$. 
In these IBM-2 calculations a bare value of $g_A=1.269$ was used and the values shown give thus a lower estimate for the half-lives. The quenching of $g_A$ is intensely discussed in literature and values from 1.2694 (the free nucleon value) to values much less than 1 have been suggested (for a review see Ref.~\cite{suh17}). For example in Ref.~\cite{pir15} quenched values from $0.6-0.8$ have been used leading to the longest half-life predictions for the mode in question \cite{kos21}.\\


Decays into the ground states are estimated to have half-lives in the range of $10^{18}-10^{26}$~a and $10^{22}-10^{24}$~a for decay modes \bbe{2} and \bbep{2}, respectively. 
The predictions of \bbe{2} transition to the first excited $0^+_1$ state vary in the range of $10^{25}-10^{29}$~a. This large range is 
partly due to the fact that the lower estimate is calculated using the bare $g_A$ value and the higher estimate is using quenched values.
For \bbep{0} the estimates are $10^{25}-10^{27}$~a assuming $m_\nu^{\rm eff} = 1$~eV.
All other listed theoretical calculations for \bbp{2} and \bbp{0} modes, as well as, for the transitions to $2^+$ states suggest half-lives longer than $10^{29}$~a. \\

 %

Previous investigations of double beta decays in \nuc{Ce}{136} and \nuc{Ce}{138} have been performed in Refs.~\cite{bel09, ber97,bel11,dan01,bel14} and were recently updated in  Ref.~\cite{bel17}. 
The main differences in this work is the use of a CeBr$_3$ crystal compared to a cerium oxide powder sample used in Ref.~\cite{bel17}, as well as the use of a fully Bayesian statistical analysis using all signature \grays\ in combined fits. 
A list of previous limits is compiled further below together with results from this work. 


\section{Experimental setup and data taking}\label{sec:setup}


\begin{table*}[ht]
\begin{center}
\caption{Radioactivity in the Al-bag\#1 and in bags from the same batch as Al-bag\#2. Decision thresholds are given at 90\% confidence level following the ISO11929:2010 standard.}
\label{tab:AlBag}
\begin{tabular}{c|cc|cc}
\hline\\[-3mm]
			& \multicolumn{2}{c|}{Al-bag\#1 }  & \multicolumn{2}{c}{Al-bag\#2} \\
radionuclide 	&  massic activity & activity per bag  & activity  & activity per bag    \\
         	 	&   [Bq/kg]                 &   [mBq]               &     [Bq/kg]       &     [mBq]    \\

\hline\\[-3mm]
\nuc{U}{238}	         &  $6.7\pm0.9$	& $190\pm26$    & $5.8\pm0.6$     & $165\pm17$\\
\nuc{Ra}{226}	         &  $< 0.04$	& $< 1.2$       & $< 0.04$        & $< 0.6$\\
\nuc{Pb}{210}	         &  $4.2\pm0.7$	& $119\pm20$    & $2.4\pm0.8$     & $68\pm23$\\
\nuc{Th}{228}	         &  $0.4\pm0.1$	& $11\pm3$      & $0.28\pm0.05$   & $8.0\pm1.5$\\
\nuc{Ra}{228}	         &  $< 0.1$	    & $< 3$         & $< 0.04$        & $< 1.2$\\
\nuc{K}{40}	             &  $< 0.3$	    & $< 9$         & $<0.1$          & $< 3$\\

\hline
\end{tabular}\\ 
\end{center}
\end{table*}

The measurements of the CeBr$_3$ crystal were carried out in the 225~m (500~m water equivalent) deep underground research facility HADES in Belgium \cite{hades, hades2}. The muon flux is reduced by about a factor 5000 compared to above ground. An HPGe-detector named ``Ge-4'' was used for all measurements of the CeBr$_3$ crystal. It is a coaxial detector with submicron top deadlayer (Canberra, XtRa-type) and has a relative efficiency of 106\% \cite{geDet}. The detector has a good long-term stability which is checked by measuring a quality control source with \nuc{Co}{60}, \nuc{Cs}{137} and \nuc{Am}{241} at regular intervals. The shift in peak centroid for the 661.7~keV \gline\ was less than 0.12~keV during the measurement campaigns. The full width at half maximum of the detector is at present 1.86~keV at 661.7~keV and 2.22~keV at 1332.5~keV. An EGSnrc \cite{egs-nrc} computer model of detector Ge-4 was established when it was installed in 2000. It has since then been refined and validated in several proficiency tests and generally produces results on an absolute scale with an accuracy of around 3\%. This computer model was used in combination with the Decay0 event generator \cite{decay0} for all calculations of all full energy peak efficiencies in this study. The detector was connected to a digital signal analyzer (Mirion’s LYNX) for high voltage and data acquisition. Spectra are collected on average every two days.\\


 A CeBr$_3$ crystal produced in March 2018 by the company Schott in Jena, Germany, was procured from the company Scionix, Netherlands \cite{scionix}. Its height and diameter are 102.55~mm and the mass is ($4380.7\pm2.0$)~g, which corresponds to \baseT{6.9456}{24} cerium atoms. Due to the hygroscopic nature of CeBr$_3$, the crystal was delivered to JRC-Geel in three layers of protective wrapping. 
The innermost protection is 0.24~mm of Teflon, enveloped by a plastic bag, and followed by an aluminized plastic bag on the outside. The aluminized bag, hereafter referred to as Al-bag, is filled with nitrogen gas. The CeBr$_3$ crystal was measured as it was delivered with these three layers of protection. After the first measurement campaign, the Al-bag used there (Al-bag\#1) was changed by Scionix and replaced by another Al-bag (Al-bag\#2).\\

The Al-bag\#1 and several Al-bags from the same batch as Al-bag\#2 were measured using \gray\ spectrometry on a low-background detector in HADES and the detected radio-impurities are reported in \tab \ref{tab:AlBag}. 
A similar Al-bag was used in a previous study of a smaller $1.5''\times1.5''$ CeBr$_3$ crystal \cite{lutter13}.\\


For practical reasons, the measurements of the CeBr$_3$ crystal were carried out in five campaigns starting May 5th 2018 and ending February 23rd 2021. 
Background and quality control sources were measured before and after each campaign. 
In total, data from 497.4 days passed the quality controls.
In the first campaign of 96.8~d the crystal was placed 25~mm above the endcap of the detector. In the later campaigns the distance was reduced to 4~mm in order to increase the \gray\ detection efficiencies. For the analysis the data is split in two datasets (M1 and M2) as indicated in \tab \ref{tab:PdMeasurementOverview} showing the resolution and detection efficiencies for both datasets.\\

\begin{figure*}[tbp]
\centerline{\includegraphics*[width=\textwidth]{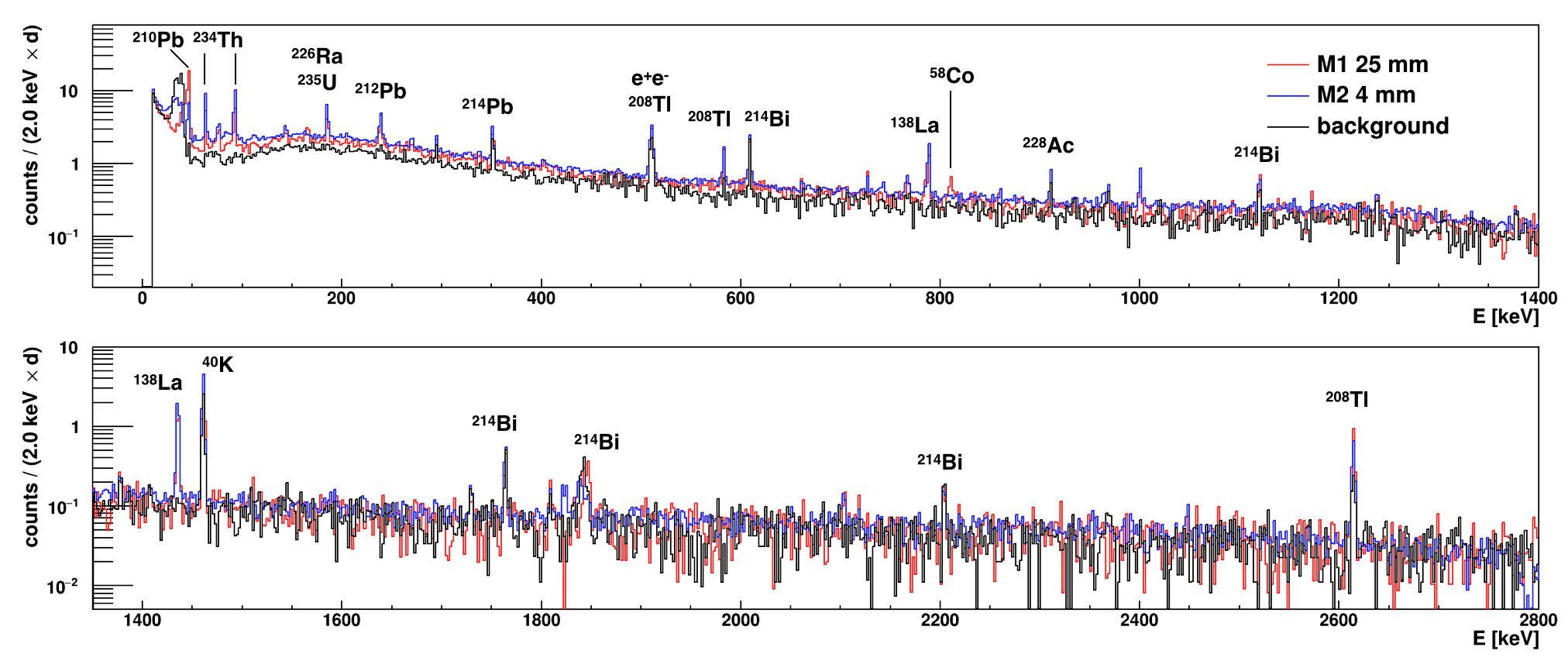}}
\caption{(color online) Recorded \gray\ spectrum of the CeBr$_3$ crystal (+Al-bag) taken in the 25~mm distance configuration (red) and in the 4~mm distance configuration (blue). Both spectra are compared to a 104~d background spectrum (black). Prominent background peaks are highlighted.}
\label{fig:SpectrumWide}
\end{figure*}

The full spectrum for measurements M1, M2 as well as a 104~d background measurement without sample is shown in \fig \ref{fig:SpectrumWide}. 
Prominent background \glines\ are labeled. 
Some of the \glines\ are only observed in the sample measurements, notably the \nuc{La}{138} \glines\ at 788.7~keV and 1435.8~keV. 
The contribution of natural decay chain background is higher in the sample measurements. This is especially visible for the low energy \glines\ of \nuc{Pb}{210} and \nuc{Th}{234} which are better shielded by the detector setup than high energy \glines\ and are thus not as prominent in the background spectrum.

\begin{table}[ht]
\begin{center}
\caption{Key parameters for the two measurement campaigns: measurement time and crystal distance to endcap. The detection efficiency ($\epsilon$) and resolution in $\sigma_E$ are shown exemplary for the emission of the 818.5~keV \gray\ in the $0^+_1$ decay mode of \nuc{Ce}{136}.}
\label{tab:PdMeasurementOverview}
\begin{tabular}{crrcc}
\hline
meas. 	&  time  & distance  & det. efficiency   & resolution $\sigma_E$      \\
\hline
M1	& 	96.8~d				&	25~mm		& $(0.44\pm0.04)$\%   & $(0.83\pm0.04)$~keV \\
M2 	&	400.6~d			&	4~mm		& $(0.65\pm0.07)$\%   & $(0.83\pm0.04)$~keV \\
\hline
\end{tabular}\\ 
\end{center}
\end{table} 

%

The radioactive impurities in the CeBr$_3$ crystal itself were quantified by subtracting the activity in the Al-bag (\tab \ref{tab:AlBag}) using Monte Carlo simulations.
They are presented in \tab \ref{tab:XtalBackground}. The most noteworthy difference between the present crystal and the crystal used in the previous measurement is the absence of \nuc{Ac}{227}, whose contribution was relatively high in the previous one.
Also, the massic activity of \nuc{La}{138} is reduced by almost a factor of 4, likely indicating a change in the production process.
A peak at 810.8~keV could be observed in the data from the first measurement campaign M1 (May 2018). The peak can be explained by the activation of copper from the shield, which is located just next to the endcap of Ge4. This copper piece had been brought above ground, end of January 2018, for a few days to be slightly modified. Refs.~\cite{Co58Argument1, Co58Argument2}  shows that production of \nuc{Co}{58} by cosmic activation is quite fast. In addition, the measured half-life ($120\pm55$)~d is compatible with the \nuc{Co}{58} half-life (70.9~d).
We also note two unidentified peaks around 1821~keV and 1825~keV in a subset of data from the M2 campaign with 4 mm detector-to-sample distance. In background data, data from the M1 campaign, and other subsets of the M2 campaign, these peaks are not observed.
An extensive search of nuclear databases and the spurious nature of the signal leads us to assume that the origin of the peaks is likely not in the CeBr$_3$ crystal.
In either case, these peaks do not interfere with the regions of interest for the double beta decay search in this work.

\begin{table*}[ht]
\begin{center}
\caption{Radioimpurities in the CeBr$_3$-crystal from both the present study and from a previous study \cite{lutter13} on a smaller crystal.}
\label{tab:XtalBackground}
\begin{tabular}{c|cc|cc}
\hline
	\multicolumn{1}{c|}{}		& \multicolumn{2}{c|}{4'' crystal 4381~g}  & \multicolumn{2}{c}{1.5'' crystal 222~g} \\
radionuclide 	&  massic activity         & activity                 & massic activity  & activity     \\
         	 	&   [mBq/kg]                 &   [mBq]                 &     [mBq/kg]      &     [mBq]    \\
\hline\\[-3mm]
\nuc{U}{238}	         &  $< 43$	        		& $< 186$	            	& $< 135$       & $< 30$\\
\nuc{Ra}{226}	         &  $< 1.1$		& $< 4.9$	            	& $< 0.5$      & $< 0.12$\\
\nuc{Pb}{210}	         & $< 400$		& $< 1700$	        	& $< 600$     & $< 134$\\
\nuc{U}{235}	         &  $< 5.4$		& $< 24$	            	& $< 1.5$      & $< 0.32$\\
\nuc{Ac}{227}	         &  $< 2.3$		& $< 10$	             	& $300\pm20$      & $64\pm4$\\
\nuc{Th}{228}	         &  $< 0.6$		& $< 2.6$	             	& $< 2$        & $< 0.44$\\
\nuc{Ra}{228}	         &  $< 0.48$		& $< 2.1$	             	& $< 0.7$         & $< 0.15$\\
\nuc{K}{40}	         &  $< 0.96$		& $< 4.2$	            	& $< 1.9$         & $< 0.43$\\
\nuc{La}{138}	         &  $2.0\pm0.2$ 	& $8.8\pm0.9$          & $7.4\pm1.0$         & $1.66\pm0.19$\\
\nuc{Ce}{139}	         &  $11.7\pm1.5$	& $51\pm7$            	& $4.3\pm0.3$         & $0.96\pm0.08$\\
\nuc{Br}{82}	         &  $5.0\pm1.0$		& $21\pm5$            	& $18\pm4$             & $3.9\pm0.9$\\
\nuc{Co}{60}	         &  $0.091\pm0.022$	& $0.4\pm0.1$          & $1.4\pm0.4$         & $0.20\pm0.12$\\

\hline
\end{tabular}\\ 
\end{center}
\end{table*}

\section{Data analysis}\label{sec:analysis}

The analysis is performed independently for each of the 35 considered decay modes using combined fits to the two datasets $d$, M1 and M2. 
Each de-excitation \gline\ $k$ in a given decay mode has its own fit region, typically $\pm10$~keV around the \gline\ of interest\footnote{Note that in some cases the fit range is adjusted to include or exclude background \glines\ on the region borders. Another exception is the $0\nu$KL $0^+_1$ decay mode where the 760.5~keV \gline\ from the de-excitation cascade and the 756.6~keV bremsstrahlung \gline\ are fitted in a single wider region.}.
 The signal count expectation $s_{d,k}$ of each \gline\ in each dataset depends on the single half-life $T_{1/2}$ parameter of the decay mode as
\begin{eqnarray}
\label{eq:CeHLtoCounts}
s_{d,k} =
\ln{2} \cdot  \frac{1}{T_{1/2}} \cdot \epsilon_{d,k} \cdot N_A \cdot t_d \cdot m \cdot f_{\rm iso}  \cdot \frac{1}{M_{\rm Ce}}\ .
\end{eqnarray}

Here, $\epsilon_{d,k}$ is the full energy detection efficiency of \gray\ $k$ in dataset $d$, $
N_A$ is Avogadro's constant,
$t_d$ is the live-time of the dataset, 
$m$ is the mass of cerium in the CeBr$_3$ crystal (1616.0~g),
$M_{\mathrm{Ce}}$ ist the molar mass of natural cerium (140.1) and
$f_{\rm iso}$ is the natural isotopic abundance of \nuc{Ce}{136} (0.186\%) and \nuc{Ce}{138} (0.251\%), respectively. 
The data is binned in 0.5~keV steps for both datasets.
The fit is performed combining all datasets and \glines\ for a given decay mode.  
The Bayesian Analysis Toolkit (BAT) \cite{Caldwell:2009kh} is used to obtain full posterior probability distributions for all parameters using Markov Chain Monte Carlo and a binned likelihood. Each free parameter in the fit has an associated prior which is either non-informative e.g.\ for the half-life, or informed by systematic uncertainties e.g.\ in case of the energy resolution. 
The likelihood $\mathcal{L}$ is defined as the product of the Poisson probabilities of each bin $i$ in the fit region for \gline\ $k$ in every dataset $d$
\begin{eqnarray}
\mathcal{L}(\mathbf{p}|\mathbf{n}) =
\prod \limits_d \prod \limits_k \prod \limits_i \frac{\lambda_{d,k,i}(\mathbf{p})^{n_{d,k,i}}}{n_{d,k,i}!} e^{-\lambda_{d,k,i}(\mathbf{p})}\ ,
\end{eqnarray}

where \textbf{n} denotes the data and \textbf{p} the set of floating parameters.
$n_{d,k,i}$ is the measured number of counts in bin $i$. $\lambda_{d,k,i}$ is the expected number of counts taken as the integral of the model $P_{d,k}$ in this bin. The model is composed of three components: 
(1) A linear background, 
(2) the Gaussian signal peak, and 
(3) a number of Gaussian background peaks. The number and type of background peaks and consequently also the number of fit parameters depend on the fit region. 
The full expression of $P_{d,k}$ is written as:
\begin{eqnarray}
P_{d,k}(E|\mathbf{p}) &&=
 B_{d,k} + C_{d,k}\left( E-E_0 \right) \label{eq:Ce_pdf}\\[2mm]
&&+  \frac{s_{d,k}}{\sqrt{2\pi}\sigma_{d,k}} 
\cdot \exp{\left(-\frac{(E-E_{k})^2}{2\sigma_{d,k}^2}\right)}\nonumber\\[2mm]
&&+ \sum \limits_{l_k} \left[\frac{b_{d,l_k}}{\sqrt{2\pi}\sigma_{d,k}} 
\cdot \exp{\left(-\frac{(E-E_{l_k})^2}{2\sigma_{d,k}^2}\right)}\right].\nonumber
\end{eqnarray}

The first row is describing the linear background with the two parameters $B_{d,k}$\ and $C_{d,k}$.
The second row is describing the signal peak with the energy resolution $\sigma_{d,k}$ and the \gline\ energy $E_k$.
The third row is describing the $l_k$ background peaks in the fit region of \gline\ $k$ with the strength of the peak $b_{d,l_k}$ and the peak position $E_{l_k} $. The same probability density function with different parameter values is used for both datasets. Hence, the same number of background peaks is used in each dataset even if not all background peaks are prominent in both datasets.\\

The free parameters $\mathbf{p}$ in the fit and their associated priors are: 
\begin{itemize}
\item 1 inverse half-life $(T_{1/2})^{-1}$ with flat prior
\item 2 x $2$ x $k$ linear background parameters $B_{d,k}$ and $C_{d,k}$ with flat priors
\item $2$ x $k$ energy resolutions $\sigma_{d,k}$ with Gaussian priors
\item $2$ x $k$ detection efficiencies $\epsilon_{d,k}$ with Gaussian priors
\item $2$ x $l_k$ x $k$ background peak strength $b_{d,l_k}$ with flat priors 
\end{itemize}

Depending on the decay mode, the contributing \glines\ (see \fig \ref{fig:Ce-136-decay}) and the background peaks in the vicinity, this amounts to 10 fit parameters in the easiest case for the \bbe{0} g.s.\ KL and LL transitions and to 79 fit parameters in the most complex case for the \bbe{0} $2_5^+$ KK, KL, and LL transitions. Note that the large number of parameters is necessary to fully describe the two datasets and multiple regions of interests within each dataset.\\

The energy resolution for each \gline\ of interest, $\sigma_{d,k}$, is obtained from calibrations and are included in the analysis with a Gaussian prior. The mean is centered around the best fit value and the width is set to an estimated systematic uncertainty of 5\%.
The detection efficiencies $\epsilon_{d,k}$ are determined with MC simulations for each dataset and decay mode and are also included with a Gaussian prior. The uncertainty is estimated at 10\%, which includes comparably negligible uncertainties from the sample mass (1\%) and the isotopic abundance (1\%).   
Additional uncertainties for the peak positions of the signal $E_k$ are typically small and neglected here. A special case is the radiative \grays\ from \bbe{0} in \nuc{Ce}{138} with $\pm5$~keV that is discussed below.\\

\begin{figure*}[tbp]
\centering
\includegraphics*[width=0.8\textwidth]{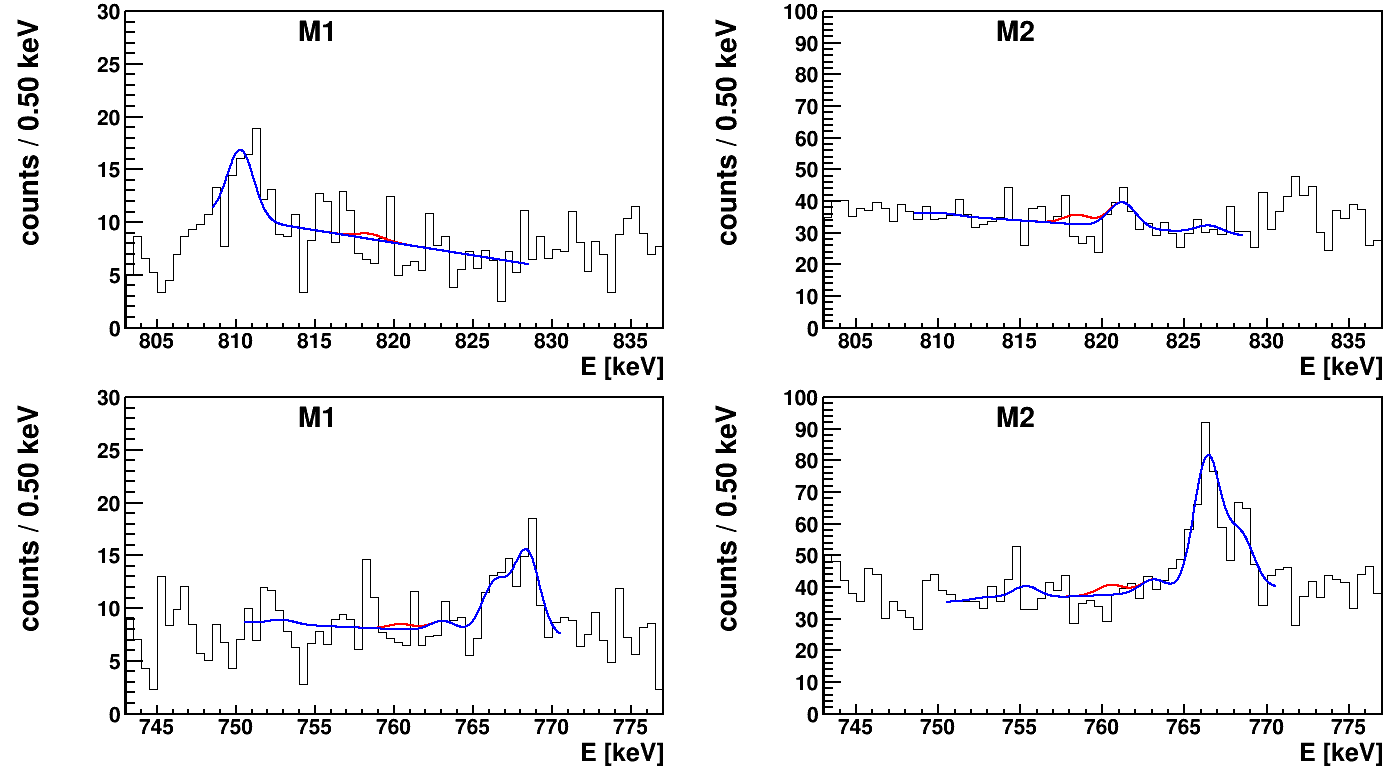}

\caption{Data and fit function for the \nuc{Ce}{136} \bbe{2} transition into the $0^+_1$ state. 
Two regions of interest for the 818.5~keV and 760.5~keV \glines\ are shown for both measurement campaigns M1 and M2. 
The blue curve shows the best fit value whereas the red curve shows the signal strength set according to the half-life limit excluded with 90\% credibility.
Background \glines\ are included in the fit as described in the text.
 }\label{fig:ExampleSpec_Ce136}
\end{figure*}

The full posterior probability obtained by BAT is used for parameter estimation. The global mode of the posterior space is the best fit. The posterior space is marginalized for $(T_{1/2})^{-1}$ and the 90\% quantile of this distribution is used for limit setting. The values quoted are 90\% credibility intervals (C.I.) on the half-life\footnote{Note that in most cases Frequentist confidence levels (C.L.) and Bayesian credibility intervals (C.I.) are numerically similar. A distinction is made due to different definitions of probability in the two concepts. In practice, numerical differences can occur in special cases when reducing multi-dimensional parameter spaces into one dimension for the parameter of interest e.g.\ $(T_{1/2})^{-1}$. This is often done by profiling in the Frequentist case and marginalization in the Bayesian case.}. 
Systematic uncertainties are naturally included via the width of Gaussian priors but typically have a small effect in the case of limit setting, where the fit is dominated by statistical uncertainties. 
Fixing the free parameters with Gaussian priors to their mean value results in about 1\% better half-life limits. \\

\fig \ref{fig:ExampleSpec_Ce136} exemplarily shows the combined fit for the \nuc{Ce}{136} \bbe{2} transition into the $0^+_1$ state for the region of interest of the 818.5~keV and 760.5~keV \glines, for both datasets, respectively. 
This fit contains 43 free parameters.
The best fit function is shown in blue and the signal strength, set to the 90\% credibility limit, is shown in red. 
The 818.5~keV fit region includes the 
810.8~keV \gline\ from \nuc{Co}{58} at 99.5\% emission probability as well as
821.2~keV (0.16\%, \nuc{Bi}{214}), and 
826.5~keV (0.12\%, \nuc{Bi}{214}) background \glines\ in the fit.
The 760.5~keV region includes the 
752.9~keV (0.13\%, \nuc{Bi}{214}),  
755.3~keV (1.0\%, \nuc{Ac}{228}), 
763.1~keV (0.64\%, \nuc{Tl}{208}), 
766.4~keV (0.32\%, \nuc{Pa}{234}), and 
768.4~keV (4.9\%, \nuc{Bi}{214}) background \glines\ in the fit.\\


Some decay modes are discussed in more detail in the following. For \bbe{0} radiative \grays\ the available energy decreases with increasing excitation level. \grays\ below 150~keV are not used in this analysis due to small detection efficiencies. Thus, these \glines\ are removed from searches for the highest excited states of the \bbe{0} $0^+_3$ and $2^+_5$ transitions.
The radiative \gray\ for the $0\nu\rm LL$ $2^+_4$ mode at 238.6~keV has the background \gline\ of \nuc{Pb}{212} (43.6\%) overlapping at the same energy and is thus ignored. 
Equally, the 287.4~keV \gline\ for the  $0\nu\rm LL$ $2^+_3$  mode is ignored due to a background \gline\ at 288.2~keV from \nuc{Bi}{214} (0.34\%)
and the 752.8 keV \gline\ in the $0\nu\rm KK$ $2^+_2$ mode due to the 752.9~keV \gline\ from \nuc{Bi}{214} (0.13\%) \\

\begin{figure*}[tbp]

\includegraphics*[width=0.4\textwidth]{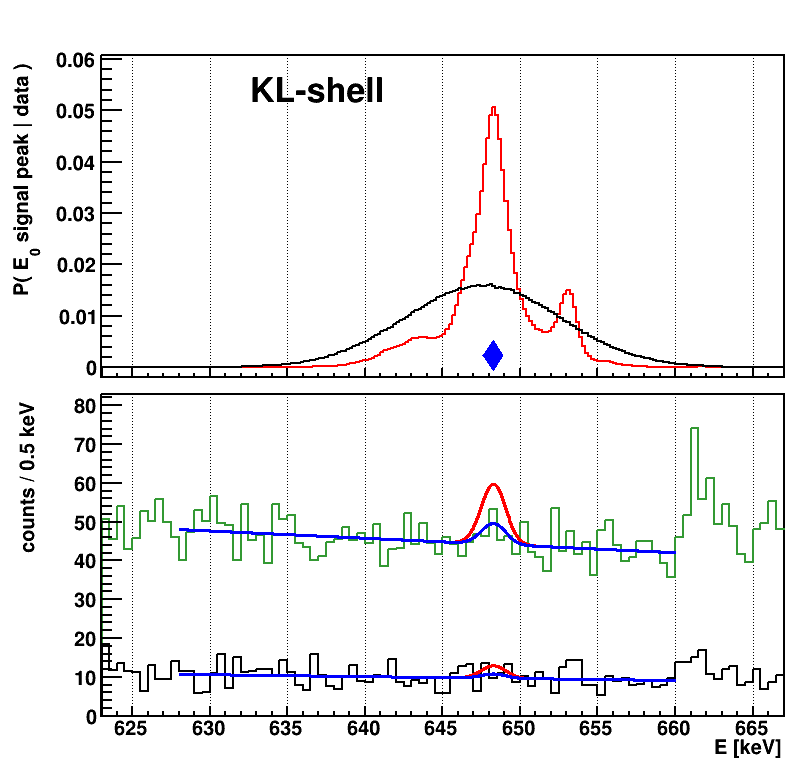}
\includegraphics*[width=0.4\textwidth]{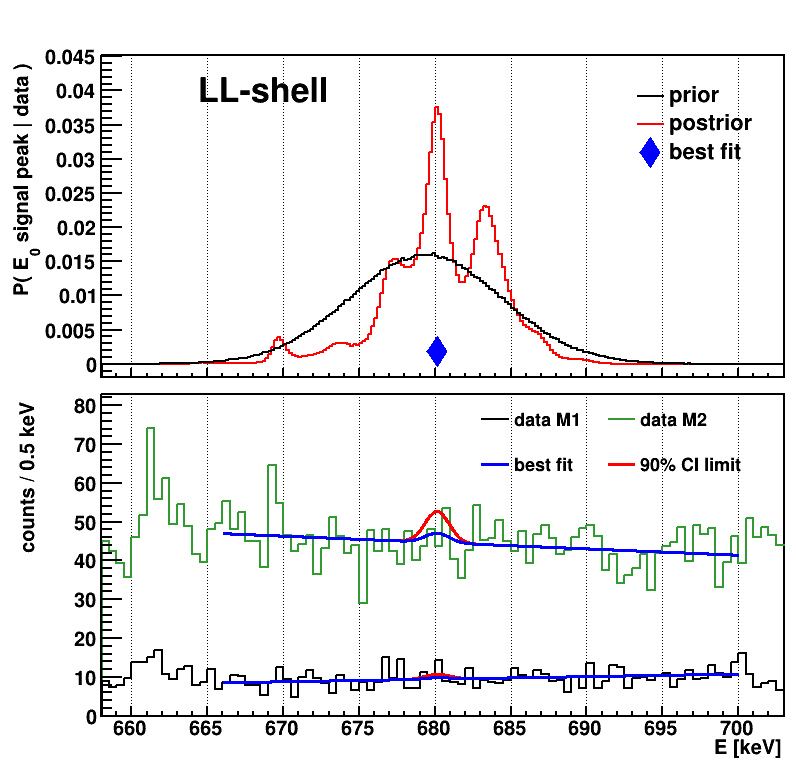}

\caption{Fit results of the \nuc{Ce}{138} KL shell capture mode (left) and LL shell capture mode (right). The top panels show the marginalized posterior probability of the signal peak energy $E_0$ in red and its input prior probability of $\pm 5$~keV in black. Also shown is the global best fit value of $E_0$ as blue diamond marker. The bottom panels show the spectra for the M1 dataset in black and the M2 dataset in green. Also shown are the fit functions for the best fit parameters in blue and the signal strength set to the 90\% credibility limit in red.
}\label{fig:Result_posteriorE0_Ce138}
\end{figure*}

For the \bbe{0} modes in \nuc{Ce}{138} the analysis becomes more complex since the Q-value and thus the signal peak energy are only known within $\pm 5$~keV. 
This is significantly larger than the energy resolution of the detector system and special care is taken to include the {\it look elsewhere effect} and correctly estimate the half-life probability within a large energy window. 
In the Bayesian framework the prior probability for the signal peak position $E_k$ (see \eq \ref{eq:Ce_pdf}) is thus included as (647.6 $\pm$ 5) keV and (679.4 $\pm$ 5)~keV for the KL and LL mode which is shown in \fig \ref{fig:Result_posteriorE0_Ce138} in the left and right top panels, respectively. The input prior probability for $E_k$ is shown in black. 
The resulting posterior probability is shown in red and shows significant structures. Any background upward fluctuation in the spectrum (shown in the bottom panels) will result in a larger or smaller probability for the signal peak at the position of the fluctuation. 
The global best fit value for $E_k$ is shown as a blue diamond marker and also in the blue best-fit functions in the bottom panels. It appears as if a non-zero signal has been found at the indicated locations of the marker. \\

\begin{figure*}[tbp]

\includegraphics*[width=0.4\textwidth]{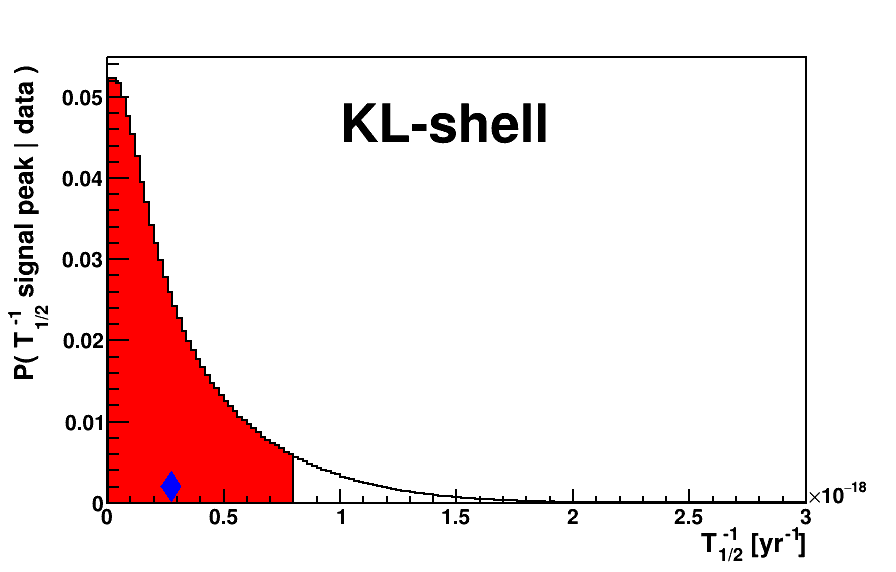}
\includegraphics*[width=0.4\textwidth]{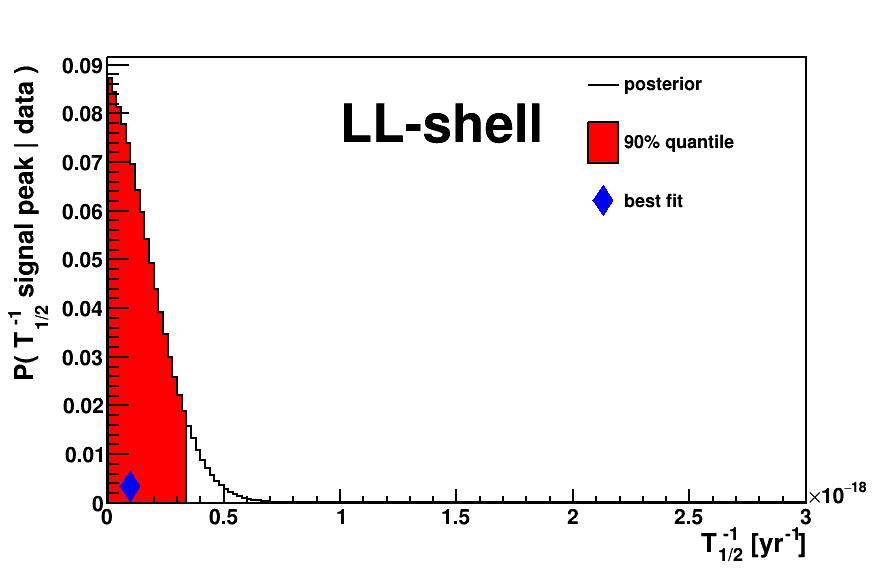}

\caption{Marginalized posterior distribution for $T_{1/2}^{-1}$ for the \nuc{Ce}{138} KL shell capture mode (left) and LL shell capture mode (right). The red shaded area shows the 90\% quantile used for limit setting. Also shown is the best fit as blue diamond marker. The global best fit is significantly different from the most likely $T_{1/2}^{-1}$ value due to the large uncertainty in the peak energy and the {\it look elsewhere effect}.  
}\label{fig:Result_posteriorT_Ce138}
\end{figure*}

However, considering the full probability for a signal anywhere in the fit region,
the most probable signal value is zero. This is shown in \fig \ref{fig:Result_posteriorT_Ce138}. The histograms show the marginalized posterior for $T_{1/2}^{-1}$ with the 90\% quantile highlighted in red and the global best fit value of $T_{1/2}^{-1}$ marked as blue diamond.
In other words, if one were to search for any peak in the wide search window, one would find one at the marked location with rather large significance as indicated by the best fit in \fig \ref{fig:Result_posteriorT_Ce138}. However, when including the {\it look elsewhere effect} and marginalizing over all possible peak positions, the evidence for a signal peak does not exist.

\section{Results}\label{results}

All fits are consistent with zero signal count and hence no double beta decay signal has been found. For 27 of the 35 considered decay modes, the $(T_{1/2})^{-1}$ posterior distributions peaks at zero. For the other eight channels a positive value is found; however zero is included within 1.1 sigma even for the most extreme case. For consistency, the 90\% quantile of the $(T_{1/2})^{-1}$ distribution is used for limit setting in all cases. The lower $T_{1/2}$ limits set in this work are shown in \tab \ref{tab:results} together with previous results. The obtained half-life limits are at a similar order of magnitude as previous results in \cite{bel17}, but improve the global constraints for most decay modes. \\

The most likely decay, \bbe{2} of \nuc{Ce}{136} to the $0^+_1$ (1579.0~keV) state in \nuc{Ba}{136}, could be constrained to \baseT{>5.0}{18}~a (90\% C.I.).\\

\begin{table}[ht]
\caption{\label{tab:results} Fit results for all accessible \nuc{Ce}{136} and \nuc{Ce}{138} decay modes. 
Previous results from Ref.\  \cite{bel17} are given as Frequentist 90\% confidence levels (C.L.) while results from this work are given as Bayesian 90\% credibility intervals (C.I.).}
\small
\begin{longtable}[c]{llll}
\hline

decay mode & final state & previous result \cite{bel17} & this work \\ 
                     & [keV] &  90\% C.L. [a] &  90\% C.I. [a] \\ 

\\[-2mm]\hline\\[-3mm]
\nuc{Ce}{136}\\
$\rm 2\nu\epsilon\epsilon$          &   $2^+_1$ (818.5)     &     \baseT{2.9}{18}     & \baseT{3.6}{18}  \\      
$\rm 2\nu\epsilon\epsilon$          &   $2^+_2$ (1551.0)    &     \baseT{3.4}{18}     & \baseT{2.5}{18}  \\      
$\rm 2\nu\epsilon\epsilon$          &   $0^+_1$ (1579.0)    &     \baseT{2.5}{18}     & \baseT{5.0}{18}  \\      
$\rm 2\nu\epsilon\epsilon$          &   $2^+_3$ (2080.0)    &     \baseT{2.8}{18}     & \baseT{2.9}{18}  \\      
$\rm 2\nu\epsilon\epsilon$          &   $2^+_4$ (2128.8)    &     \baseT{1.4}{18}     & \baseT{2.6}{18}  \\      
$\rm 2\nu\epsilon\epsilon$          &   $0^+_2$ (2141.3)    &     \baseT{4.4}{18}     & \baseT{5.9}{18}  \\      
$\rm 2\nu\epsilon\epsilon$          &   $(2)^+_5$ (2222.7)  &     \baseT{2.0}{18}     & \baseT{3.6}{18}  \\      
$\rm 2\nu\epsilon\epsilon$          &   $0^+_3$ (2315.3)    &     \baseT{2.5}{18}     & \baseT{4.4}{18}  \\      

$\rm 0\nu KL$                       &   $0^+_0$ (g.s.)      &     \baseT{3.4}{18}     & \baseT{7.3}{18}  \\      
$\rm 0\nu LL$                       &   $0^+_0$ (g.s.)      &     \baseT{8.4}{18}     & \baseT{4.1}{18}  \\      

$\rm 0\nu KK$                       &   $2^+_1$ (818.5)     &     \baseT{3.0}{18}     & \baseT{7.7}{18}  \\      
$\rm 0\nu KL$                       &   $2^+_1$ (818.5)     &     \baseT{3.0}{18}     & \baseT{5.6}{18}  \\      
$\rm 0\nu LL$                       &   $2^+_1$ (818.5)     &     \baseT{3.0}{18}     & \baseT{3.5}{18}  \\      

$\rm 0\nu KK$                       &   $2^+_2$ (1551.0)    &     \baseT{2.9}{18}     & \baseT{2.4}{18}  \\      
$\rm 0\nu KL$                       &   $2^+_2$ (1551.0)    &     \baseT{2.9}{18}     & \baseT{2.2}{18}  \\      
$\rm 0\nu LL$                       &   $2^+_2$ (1551.0)    &     \baseT{2.9}{18}     & \baseT{2.0}{18}  \\      

$\rm 0\nu KL$                       &   $0^+_1$ (1579.0)    &     \baseT{2.2}{18}     & \baseT{4.6}{18}  \\      
$\rm 0\nu LL$                       &   $0^+_1$ (1579.0)    &     \baseT{2.2}{18}     & \baseT{4.4}{18}  \\      

$\rm 0\nu KK$                       &   $2^+_3$ (2080.0)    &     \baseT{2.6}{18}     & \baseT{2.5}{18}  \\     
$\rm 0\nu KL$                       &   $2^+_3$ (2080.0)    &     \baseT{2.6}{18}     & \baseT{1.8}{18}  \\      
$\rm 0\nu LL$                       &   $2^+_3$ (2080.0)    &     \baseT{2.6}{18}     & \baseT{2.7}{18}  \\      

$\rm 0\nu KK$                       &   $2^+_4$ (2128.8)    &     \baseT{1.6}{18}     & \baseT{3.1}{18}  \\      
$\rm 0\nu KL$                       &   $2^+_4$ (2128.8)    &     \baseT{1.6}{18}     & \baseT{2.2}{18}  \\      
$\rm 0\nu LL$                       &   $2^+_4$ (2128.8)    &     \baseT{1.6}{18}     & \baseT{2.5}{18}  \\      

$\rm 0\nu KL$                       &   $0^+_2$ (2141.3)    &     \baseT{4.2}{18}     & \baseT{6.3}{18}  \\      
$\rm 0\nu LL$                       &   $0^+_2$ (2141.3)    &     \baseT{4.2}{18}     & \baseT{6.2}{18}  \\      

$\rm 0\nu KK$                       &   $(2)^+_5$ (2222.7)  &     \baseT{2.0}{18}     & \baseT{3.6}{18}  \\      
$\rm 0\nu KL$                       &   $(2)^+_5$ (2222.7)  &     \baseT{2.0}{18}     & \baseT{3.5}{18}  \\      
$\rm 0\nu LL$                       &   $(2)^+_5$ (2222.7)  &     \baseT{2.0}{18}     & \baseT{3.5}{18}  \\      

$\rm 0\nu KL$                       &   $0^+_3$ (2315.3)    &     \baseT{2.5}{18}     & \baseT{4.7}{18}  \\      
$\rm 0\nu LL$                       &   $0^+_3$ (2315.3)    &     \baseT{2.5}{18}     & \baseT{4.6}{18}  \\      

$\rm 0\nu\epsilon\beta^+$           &   $2^+_1$ (818.5)     &     \baseT{2.3}{18}     & \baseT{3.0}{18}  \\     
$\rm 2\nu\epsilon\beta^+$           &   $2^+_1$ (818.5)     &     \baseT{2.4}{18}     & \baseT{3.0}{18}  \\      

\\[-2mm]\hline\\[-3mm]

\nuc{Ce}{138}\\                                 
$\rm 0\nu KL$                       &   $0^+_0$ (g.s.)      &     \baseT{8.3}{17}     & \baseT{1.3}{18}  \\      
$\rm 0\nu LL$                       &   $0^+_0$ (g.s.)      &     \baseT{4.2}{18}     & \baseT{3.1}{18}  \\

\hline
\end{longtable}
\end{table}

This analysis considers many decay modes - some with a complex decay scheme and many de-excitation \grays.
The order of magnitude of the half-life limits is mainly determined by the exposure and background of the experiment. 
Smaller variations are due to different detection efficiencies of \grays\ in the different decay modes. 
The use of information from all prominent \glines\ for a given decay mode in a combined fit largely mitigates strong differences as would be seen  when e.g.\ only a single \gline\ is used for limit setting. 
Another source of variations are statistical fluctuations of the background. Also here, the use of a combined fit to multiple \glines\ reduces the probability of strong background fluctuation compared to using a single \gline\ fit. 
It is also worth noting that the approach of combined fits reduces the selection bias. This would occur, if one selects the limit from the \gline\, which happens to result in the highest limit.
Especially for decay modes with multiple \grays\ at similar detection efficiency, this would systematically select favorable background fluctuations.\\    

Each decay mode is analyzed independently. However, the value of the limits is not independent, since many decay modes share the same \gline. This is especially true for the 818.5~keV \gline\ from the \nuc{Ce}{136} $2^+_1$ state which is the final \gray\ in the de-excitation cascade.

\section{Discussion and Conclusion}

A 4381~g CeBr$_3$ crystal was measured with \gray\ spectrometry over the last three years at the HADES underground lab.
An extensive search for double electron capture transitions in cerium isotopes has been performed on 479.4~days of data. No signals have been observed and 90\% credibility limits have been set using a Bayesian analysis for all accessible decay modes in \nuc{Ce}{136} and \nuc{Ce}{138}. 
Previously existing limits could be improved by up to a factor of two. Special care has been taken to avoid selection biases and {\it look elsewhere effects} using all available spectral information in the analysis as well as implementing knowledge of systematic uncertainties using Bayesian priors.\\  

The measurement was performed in a state-of-the-art low-background \gray\ spectrometry setup for an extended time. 
Further reducing the radioactive background or increasing the measurement time / exposure is limited in feasibility and practicality. 
Instead, a much stronger improvement on the half-life sensitivity may be achieved by instrumenting the CeBr$_3$ crystal as a scintillation detector - its intended purpose. 
This allows analysis-driven background rejection by coincidence or anti-coincidence requirements between the CeBr$_3$ and the HPGe detectors. 
In addition, turning the crystal into a detector would enable searches in the source$=$detector configuration with a two order of magnitude increase in detection efficiency close to 100\%. Furthermore, additional decay modes become accessible, e.g.\ those which only emit X-rays, which do not easily escape the crystal, or \nuc{Ce}{142} \bb{0/2} modes with continuous double beta spectrum. 
With enough resources, isotopic enrichment of the low mass side of cerium isotopes can increase the half-life sensitivity by multiple orders of magnitude.  Natural abundances of  \nuc{Ce}{138} and \nuc{Ce}{136} are only 0.25\% and 0.19\%, respectively, with strong potential for enrichment.\\

The newly achieved experimental constraints on the half-lives of the order of \baseTsolo{18-19}~a are still considerably far away from theoretical predictions. 
The shortest half-lives are expected for the \bbe{2} ground state transitions which were not accessible in this search. However, they are predicted as low as \baseT{3}{18}~a for \nuc{Ce}{136} and come into reach with an instrumented CeBr$_3$ crystal.
\bbep{2} decay modes to the ground state have slightly longer half-life predictions but an enhanced experimental signature with two annihilation photons. This signature could not be exploited in this setup but becomes a powerful discriminator using CeBr$_3$-HPGe coincidences.
The shortest decay with \gray\ emission is the \nuc{Ce}{136} $0^+_1$ mode, predicted at \baseT{5.7}{25}~a. Reaching this sensitivity requires significant improvement and investment. However, the only two observed double beta decay excited state transitions in \nuc{Mo}{100} and \nuc{Nd}{150} were measured at a significant lower half-life than predicted \cite{ESReview}. Surprises are possible for the complicated nuclear physics of double beta decay and future experimental searches are well motivated.\\

%
%
%

\section{Acknowledgements}

This work received support from the EC-JRC open access scheme EUFRAT under Horizon-2020, project number 35375-1. 
EIG EURIDICE and the staff of HADES are gratefully acknowledged for their work.
One of the authors (A.\ O.) acknowledges the support from the Extreme Light Infrastructure Nuclear Physics (ELI-NP) Phase II, a project co-financed by the Romanian Government and the European Union through the European Regional Development Fund, the Competitiveness Operational Programme (1/07.07.2016, COP, ID 1334).
Another authors (J.\ K.) acknowledges the support from the Academy of Finland, Grant Nos. 314733 and 320062.


\end{document}